\documentclass[12pt,nofootinbib,amsmath,amssymb,preprint,showpacs]{revtex4}
\usepackage[T1]{fontenc} \usepackage[latin1]{inputenc}
\usepackage{amsmath,color} \usepackage[dvips]{graphicx}
\usepackage{epsfig} \usepackage{amssymb}

\begin{document}

\title{Dynamical Eigenmodes of a Polymerized Membrane}

\author{Rick Keesman}\affiliation{Institute for Theoretical Physics,
  Universiteit Utrecht, Leuvenlaan 4, 3584 CE Utrecht, The
  Netherlands} \author{Gerard T. Barkema} \affiliation{Institute for
  Theoretical Physics, Universiteit Utrecht, Leuvenlaan 4, 3584 CE
  Utrecht, The Netherlands}\affiliation{Instituut-Lorentz,
  Universiteit Leiden, Niels Bohrweg 2, 2333 CA Leiden, The
  Netherlands} \author{Debabrata Panja} \affiliation{Institute for
  Theoretical Physics, Universiteit Utrecht, Leuvenlaan 4, 3584 CE
  Utrecht, The Netherlands}\affiliation{Institute of Physics,
  Universiteit van Amsterdam, Postbus 94485, 1090 GL Amsterdam, The
  Netherlands}

\begin{abstract}
We study the bead-spring model for a polymerized phantom membrane in the
overdamped limit, which is the two-dimensional generalization of the
well-known Rouse model for polymers. We derive the {\it exact\/}
eigenmodes of the membrane dynamics (the ``Rouse modes''). This allows
us to obtain exact analytical expressions for virtually any
equilibrium or dynamical quantity for the membrane. As examples we
determine the radius of gyration, the mean square displacement of a
tagged bead, and the autocorrelation function of the difference vector
between two tagged beads. Interestingly, even in the presence of
tensile forces of any magnitude the Rouse modes remain the exact
eigenmodes for the membrane.  With stronger forces the membrane
becomes essentially flat, and does not get the opportunity to
intersect itself; in such a situation our analysis provides a useful
and exactly soluble approach to the dynamics for a realistic model
flat membrane under tension.
\end{abstract}


\maketitle

\section{Introduction\label{sec1}}

Bead-spring models for linear polymers, wherein the polymer is
modelled as a sequence of beads connected by springs, play a central
role in the theory of polymer dynamics. For a linear bead-spring
polymer chain, with position ${\bf R}_n$ of the $n$-th bead, $n=1\dots
N$, the potential energy is given by $\displaystyle{ U= \frac{k}{2}
  \sum_{n=1}^{N-1} \left({\bf R}_n - {\bf R}_{n+1} \right)^2}$, in
which $k$ is the spring constant. Within the scope of bead-spring
models, the Rouse model \cite{rouse} holds a special status, as it
provides a framework for linear polymers by which virtually any static
and dynamical quantity of interest can be calculated analytically
\cite{doi,de}. In the Rouse model the dynamics of the beads is
formulated in the overdamped limit. For solving the Rouse model one
considers the so-called Rouse modes (rather than the equations of
motion for the individual beads). Using the mode amplitude correlation
functions, the quantities of interest for a phantom chain can be
analytically tracked by reconstructing them from the modes
\cite{doi,de}. The Rouse model has been shown useful for self-avoiding
linear polymers \cite{chain1}, and to show that the dynamics of a
tagged bead in a linear bead-spring model is described by the
Generalized Langevin Equation (GLE) \cite{jstat1,jstat2}.

It is therefore a natural question to ask if the analytical solution
of the Rouse model can be generalized to higher dimensions, e.g.,
making it useful for the dynamics of polymerized membranes. There is a
considerable amount interest in the properties of polymerized
membranes \cite{book}. The interest stems not only from the point of
view of a fundamental understanding, but also because of their
importance in biology and chemistry. Instances of polymerized
membranes with a fixed connectivity can be found in 2D-cytoskeletons
of cells \cite{cells1,cells2,cells3} and graphite oxide sheets
\cite{sheet1,sheet2,sheet3}.

Static properties of polymerized membranes as a two-dimensional
extension of (one-dimensional) linear polymers came into fashion in
the late 80s. At that time the driving question centered on the
equilibrium properties of polymerized membranes, in particular, the
scaling of the radius of gyration of a polymerized membrane with its
lateral size, while the corresponding scaling behavior for phantom and
self-avoiding linear polymers were already well-known. Sophisticated
renormalized group approaches were developed to this end
\cite{renorm1,renorm2,renorm3,renorm4}, accompanied by
computer simulations
\cite{mcsim1,mcsim2,mcsim3,mcsim4,mcsim5,mcsim6,mcsim7,mcsim8,mcsim9}. A
notable outcome of these studies is that the radius of gyration $R_g$
of a membrane scales with its lateral size $N$ as $R_g\sim\log N$ for
phantom, and as $R_g\sim N$ for self-avoiding membranes.

In contrast to the equilibrium properties, the dynamical properties of
membranes have been studied with less intensity. Apart from scaling
analyses \cite{dynamana1,dynamana2,dynamana3}, the bulk of the
research on the dynamics of polymerized membranes are heavily
dominated by computer simulations
\cite{dynammcsim1,dynammcsim2,dynammcsim3,dynammcsim4}, leaving exact
analytical results on the dynamics of membranes a relatively open
area. In this paper we attempt to fill this void --- we consider
two-dimensional square-polymerized phantom membranes embedded in a
three-dimensional space, both in the absence and the presence of
tensile forces --- and perform a Rouse mode analysis. We show that the
Rouse modes are the {\it exact\/} eigenmodes of the membrane
dynamics. Akin to the Rouse model for bead-spring linear polymers, our
exercise allows us to exactly solve for the static and dynamic
properties of phantom membranes; in this process also deriving the
exact expression for the scaling of $R_g$ as a function of the
membrane's lateral size $N$ (which confirms the $R_g\sim\log N$
scaling), as well as show increased fluctuations at the edge of the
membrane \cite{edge1,edge2}.  In particular, we draw the reader's
attention to the exact solution under tensile forces, for which forces
the membrane essentially encounters a flat geometry. Unlike a phantom
membrane in equilibrium, in such a situation the membrane becomes
essentially flat, does not get the opportunity to intersect itself,
and therefore provides a highly useful and exactly soluble approach to
the dynamics for a realistic model flat membrane. We also note that
the methods are generalizable to arbitrary internal and spatial
dimensions.

The structure of this paper is as follows. In Sec. \ref{sec2} we
describe the dynamical equations without forces, and their
diagonalization by the Rouse modes. In Sec. \ref{sec3} we derive the
scaling of the radius of gyration, the mean-square displacement of a
tagged bead and the autocorrelation function of a vector connecting
two beads. In Sec. \ref{sec4} we incorporate tensile forces in the
dynamical equations and derive the new Rouse modes that diagonalize
them. We finally end the paper with a discussion in Sec. \ref{sec5}.

\section{Diagonalization of equations of motion for the polymerized
  membrane and the Rouse modes}\label{sec2}

We consider a rectangular polymerized membrane, for which $N=L_1
\times L_2$ beads are connected in a perpendicular toplogy, with ${\bf
  R}_{{\bf n}}(t)$ denoting the spacial position at time $t$ of the
bead internally labeled as ${\bf n}$. Naturally for a membrane it is
convenient to take two numbers ${\bf n}=(n_1,n_2)$ for labeling the
beads with $n_i=1,\dots,L_i$. The potential energy for a
square-polymerized membrane is a simple extension of the Hamiltonian
for a linear polymer in two internal dimensions
\begin{equation}
U =   \frac{k}{2} \sum_{n_1,n_2=1}^{L_1-1,L_2} \left( {\bf
  R}_{n_1,n_2} - {\bf R}_{n_1+1,n_2} \right)^2+ \frac{k}{2}
\sum_{n_1,n_2=1}^{L_1,L_2-1} \left( {\bf R}_{n_1,n_2} - {\bf
  R}_{n_1,n_2+1} \right)^2
\end{equation}
for some spring constant $k$.

In the absence of externally applied forces, in the overdamped limit
the dynamics of each bead of the membrane is described by
\begin{equation}
\label{eq:eommon}
\frac{d {\bf R}_{{\bf n}}}{dt}= - \frac{1}{\zeta} \frac{\partial
  U}{\partial {\bf R}_{{\bf n}}} + {\bf g}_{{\bf n}},
\end{equation}
where $\zeta$ is the friction coefficient of the solvent and ${\bf
  g}_{{\bf n}}$ is the thermal force on the ${\bf n}$-th bead. The
thermal forces are uncorrelated between different beads, as well as in
time, i.e.,
\begin{equation}
\label{eq:thermal}
\left< {\bf  g}_{{\bf m}}(t) \cdot {\bf  g}_{{\bf n}}(t^\prime)
\right> = \frac{6 k_B T}{\zeta} \delta_{{\bf m}{\bf
    n}}\delta(t-t^\prime),
\end{equation}
with Boltzmann constant $k_B$ and temperature $T$.

For a linear polymer with positions ${\bf R}_{n}(t)$ of beads
$n=1,\dots,N$ at time $t$ the Rouse modes are given by \cite{doi,de}
\begin{equation}
{\bf X}_{p}(t)=\frac{1}{N} \sum_{n=1}^{N} \cos \left[
  \frac{\pi(n-1/2)p}{N} \right] {\bf R}_{n}(t),
\label{modedef}
\end{equation}
with $p=0,\dots,N-1$ and the inverse given by
\begin{equation}
{\bf R}_{n}(t)= {\bf X}_{0}(t)+ 2\sum_{p=1}^{N-1} \cos \left[
  \frac{\pi(n-1/2)p}{N} \right] {\bf X}_{p}(t)
\label{invmodedef}
\end{equation}
The Rouse modes for the membrane are quite similar to those of the
linear chain: since topologically the internal directions
(connectivity) are orthogonal, the Rouse modes will be products of
Rouse modes for a linear polymer. Below we introduce the following
definitions for an elegant notation and proof of the independence of
the Rouse modes.

First we define
\begin{equation}\label{eq:betadef}
\beta_{p_i} = \left\{
	\begin{array}{ll}
        2 & \mbox{if } p_i = 1,\dots,L_i-1\\ 1 & \mbox{if } p_i = 0
	\end{array}
\right. , \quad \beta_{{\bf p}} = \beta_{p_1}\beta_{p_2}
\end{equation}
and
\begin{equation}
f_{p_i}(n_i) = \cos \left[ \frac{\pi p_i (n_i -1/2)}{L_i} \right]  ,
\quad f_{{\bf p}}({\bf n}) = f_{p_1}(n_1)f_{p_2}(n_2)
\label{eq:fdef}
\end{equation}
The orthogonality relation at the basis of the proof for the Rouse
modes for a linear polymer chain can then be generalized for the
membrane
\begin{equation}
\frac{1}{N} \sum_{{\bf n}} \beta_{{\bf p}} f_{{\bf p}}({\bf n})
f_{{\bf q}}({\bf n})= \delta_{{\bf p}{\bf q}},
\label{eq:ortho2}
\end{equation}
where the summation is taken over all allowed values of ${\bf n}$.
The Rouse modes amplitudes and the corresponding inverse are then
given by
\begin{equation}
{\bf X}_{{\bf p}}(t) = \frac{1}{N} \sum_{{\bf n}} f_{{\bf p}}({\bf n})
{\bf R}_{{\bf n}}(t),
\label{eq:eigenmode}
\end{equation}
\begin{equation}
{\bf R}_{{\bf n}}(t) = \sum_{{\bf p}} \beta_{{\bf p}} f_{{\bf p}}({\bf
  n}) {\bf X}_{{\bf p}}(t).
\label{eq:inverse}
\end{equation}
The position of the center-of-mass ${\bf R}_{\text{cm}}(t)$ at time
$t$ simply equals ${\bf X}_{{\bf 0}}(t)$.

Next, we note that when the equations of motion from
Eq. (\ref{eq:eommon}) for the beads are expressed one gets a term for
each internal dimension equal to that of a term for a bead in a linear
polymer. Taking the time derivative in Eq. (\ref{eq:eigenmode}) and
plugging in the equations of motion for the beads and then the inverse
we get
\begin{equation}
\frac{d {\bf X}_{{\bf p}}(t)}{d t} = - \alpha_{{\bf p}} {\bf X}_{{\bf
    p}}(t) + {\bf G}_{{\bf p}}(t)
\label{eq:eomeigenmode3}
\end{equation}
\begin{equation}
\mbox{ with } \alpha_{{\bf p}} \equiv 4\frac{k}{\zeta} \left(
\sin^2\left[ \frac{\pi p_1}{2 L_1} \right] + \sin^2\left[ \frac{\pi
    p_2}{2 L_2} \right] \right) \mbox{ and } {\bf G}_{{\bf p}}(t) =
\frac{1}{N} \sum_{{\bf n}} f_{{\bf p}}({\bf n}) {\bf g}_{{\bf n}}(t),
\label{alphap}
\end{equation}
i.e., the Rouse modes diagonalizes the equations of motion. This set
of linearized differential equations can be solved just as for a
linear polymer by using the correlation function for the thermal
forces given by Eq. (\ref{eq:thermal}). By using the orthogonality
relation in Eq. (\ref{eq:ortho2})
\begin{equation}
\left< {\bf G}_{{\bf p}}(t) \cdot {\bf G}_{{\bf q}}(t^\prime) \right>
= \frac{6 k_B T}{\zeta N \beta_{{\bf p}}} \delta_{{\bf p}{\bf
    q}}\delta(t-t^\prime),
\label{eq:thermal2}
\end{equation}
so that the following two-point correlation functions for the
eigenmodes can be derived:
\begin{align}
&X_{\bf 0 0}(t) \equiv \langle [ {\bf X}_{{\bf 0}}(t) - {\bf X}_{{\bf
        0}}(0) ]^2 \rangle \equiv \langle [ {\bf R}_{\text{cm}}(t)
    -{\bf R}_{\text{cm}}(0) ]^2 \rangle = \frac{6 k_B T}{\zeta N} t
\label{eq:macfb} \\&X_{{\bf p} {\bf q}}(t) \equiv \langle {\bf X}_{{\bf
p}}(t) \cdot {\bf X}_{{\bf q}}(0) \rangle = \frac{3 k_B T}{\zeta N
  \beta_{{\bf p}}}\frac{1}{\alpha_{{\bf p}}} \exp \left[-\alpha_{{\bf
      p}} t\right] \delta_{{\bf p}{\bf q}} \quad \mbox{with } {\bf p}
\neq {\bf 0} .
\label{eq:macfa}
\end{align}
Just like the case of a linear polymer, any correlation function of
interest for the membrane can be derived using
Eqs. (\ref{eq:eomeigenmode3}-\ref{eq:thermal2}). We address a few of
them in the following section.

\section{A few properties of interest for a membrane\label{sec3}}

All throughout this section, for the sake of simplicity, we restrict
ourselves to square membranes with $L_1=L_2=L$, and thus $N=L\times
L$.

\subsection{Scaling of the radius of gyration $R_g$}

\begin{figure}[ht] \centering
\epsfig{file=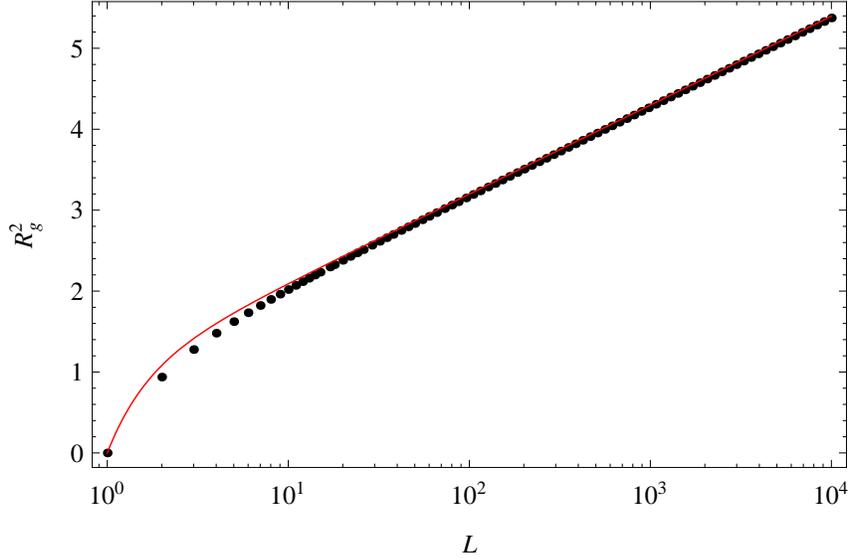,width=0.7\linewidth,clip=}
\caption{The radius of gyration squared $R_g^2$ for a square membrane
  as a function of $N=L \times L$ beads, with $k_BT/k$ set to
  unity. The black dots represents the sum in Eq. (\ref{eq:rog1})
  evaluated exactly for $L$ up to $L=10^4$, while the continuous red
  line represents the function $(N-1)/N+3/(4\pi)\log N$.}
    \label{fig:rog}
\end{figure}

The radius of gyration squared for the membrane is defined by
\begin{align}
R^2_g \equiv& \frac{1}{N} \sum_{{\bf n}} \left\langle \left[{\bf
    R}_{{\bf n}}(t)-{\bf R}_{\text{cm}}(t)
  \right]^2\right\rangle\nonumber\\ =&\frac{1}{N}\sum_{{\bf n}}
\sum_{{\bf p}\neq{\bf 0}} \sum_{{\bf q}\neq{\bf 0}} \langle {\bf
  X}_{{\bf p}}(t) \cdot {\bf X}_{{\bf q}}(t) \rangle \beta_{{\bf
    p}}\beta_{{\bf q}} f_{{\bf p}}({\bf n}) f_{{\bf q}}({\bf n}),
\label{rgsquared}
\end{align}
in which Eq. (\ref{eq:inverse}) is used for the second line. Having
plugged in Eqs. (\ref{eq:ortho2}) and (\ref{eq:macfa}) the squared
radius of gyration reduces to
\begin{align}
R^2_g = \frac{3 k_B T}{\zeta N} \sum_{{\bf p}\neq{\bf 0}}
\frac{1}{\alpha_{{\bf p}}}.
\label{eq:rog1}
\end{align}
The summation (\ref{eq:rog1}), using the definition of $\alpha_{\bf
  p}$ (\ref{alphap}), can be split into two parts:
\begin{align}
R^2_g = \frac{k_B T}{k} \left( \frac{3}{2 N} \sum_{p=1}^{L-1}
\sin^{-2}\left[\frac{\pi p}{2 L}\right] + \frac{3 k}{\zeta N}
\sum_{p_1=1}^{L-1} \sum_{p_2=1}^{L-1} \frac{1}{\alpha_{{\bf
      p}}}\right) = \frac{k_B T}{k} \left( \mathcal{I}+\mathcal{J}
\right).
\label{iplusj}
\end{align}
The first term in Eq. (\ref{iplusj}) can be evaluated exactly:
$\mathcal{I} = (N-1)/N$. For the second term we take the long polymer
limit $N\gg 1$ so that $\alpha_{{\bf p}}$ can be expanded and the sum
approximated by an integral with $x_i = p_i/L$  where the integral is
over a unit square in the positive quadrant excluding the small area
near the origin. The second approximation we make is replacing the
area of integration by that of the positive quadrant of a unit circle
excluding the small area near the origin which can be solved
analytically
\begin{align}
\mathcal{J} = \frac{3}{\pi^2} \int \frac{d{\bf x} }{x^2} = \frac{3}{2
  \pi} \int_{1/L}^{1} \frac{dr }{r} = \frac{3}{4 \pi} \log N.
\end{align}
The radius of gyration squared for a square membrane in the long
polymer limit thus becomes
\begin{align}
R^2_g = \frac{k_B T}{k} \left(\frac{N-1}{N} + \frac{3}{4 \pi} \log N
\right).
\label{rgscaling}
\end{align}
A comparison between the two terms in Eq. (\ref{rgscaling}) shows that
the second one is bigger than the first one for $N>61$. In
Fig. \ref{fig:rog} we present a comparison between the analytical
result (\ref{rgscaling}) and the exact evaluation of
Eq. (\ref{eq:rog1}) for $L$ up to $L=10^4$. Our result
(\ref{rgscaling}) confirms the $\log N$ scaling of $R_g^2$ obtained
earlier by field-theory methods\cite{renorm1,renorm2,renorm3,renorm4}.

\subsection{Mean-square displacement of a tagged bead}

Let us define $\Delta{\bf R}_{{\bf n}}(t) \equiv {\bf R}_{{\bf
    n}}(t)-{\bf R}_{{\bf n}}(0)$ for the ${\bf n}$-th bead. We use
Eq. (\ref{eq:inverse}) to write
\begin{align}
\langle \Delta{\bf R}_{{\bf n}}(t)^2 \rangle = &\langle [ {\bf
    X}_{{\bf 0}}(t) - {\bf X}_{{\bf 0}}(0) ]^2 \rangle \nonumber\\ &+2
\sum_{{\bf p} \neq {\bf 0}} \sum_{{\bf q} \neq {\bf 0}} \beta_{{\bf
    p}} \beta_{{\bf q}} f_{{\bf p}}({\bf n}) f_{{\bf q}}({\bf n})
\left\{ X_{{\bf p} {\bf q}}(0) - X_{{\bf p} {\bf q}}(t)\right\}.
\label{eq:msd1}
\end{align}
Using Eqs. (\ref{eq:macfb}-\ref{eq:macfa}) this simplifies to
\begin{equation}
\langle \Delta{\bf R}_{{\bf n}}(t)^2 \rangle = \frac{6 k_B T}{\zeta N}
t + \frac{6 k_B T}{\zeta N} \sum_{{\bf p} \neq {\bf 0}}
\frac{\beta_{{\bf p}}}{\alpha_{{\bf p}}} f_{{\bf p}}({\bf n})^2
\left\{1 -e^{-\alpha_{{\bf p}} t}\right\}.
\label{eq:msd2}
\end{equation}
At short times $k t/\zeta\ll1/8$ the exponential in the second term
can be expanded, and the sum is exactly evaluated. In comparison to
the second term the first one can be neglected to obtain
\begin{equation}
\langle \Delta{\bf R}_{{\bf n}}(t)^2 \rangle = \frac{6 k_B T}{\zeta} t
\frac{1}{N} \sum_{{\bf p} \neq {\bf 0}} \beta_{{\bf p}} f_{{\bf
    p}}({\bf n})^2 = \frac{6 k_B T}{\zeta} t,
\label{eq:msd3}
\end{equation}
which agrees with the mean-square displacement of a free bead, and
confirms the wisdom that at short times the beads do not feel the
connectivity.

At very long times the exponentials in Eq. (\ref{eq:msd2}) reduce to
zero, the second term becomes a fixed number independent of $t$ [but
  dependent on the location of the bead: the sum is largest when the
  bead is located at the corner of the membrane, and equals $6 k_B
  T\left(\log N/\pi - \pi/4 + 2/3\right)/k$]; i.e., for
$kt/\zeta\gg N\log N/\pi$ the first term of Eq. (\ref{eq:msd1})
dominates, and the motion of the tagged bead becomes simply
diffusive, with the same diffusion coefficient as that of the
center-of-mass.

At intermediate times the mean-square displacement of the tagged bead
will depend on its location on the membrane. We work out three
difference cases when the tagged bead is (a) the central bead of the
membrane, (b) located in the middle of an edge, and (c) a corner bead.
The enhanced mobility of edge and corner beads was already reported
for self-avoiding membranes \cite{edge1,edge2}.

(a) For the mean-square displacement of a bead at the center of the
membrane $\langle \Delta{\bf R}_{\text{m}}(t)^2 \rangle$ the sum in
Eq. (\ref{eq:msd2}) will only be over even values of $p_i$. In the
long polymer limit the sum can be converted to an integral with
$x_i=p_i/L$, such that
\begin{equation}\langle \Delta{\bf R}_{\text{m}}(t)^2
\rangle = \frac{6 k_B T}{\zeta} \int_0^1 dx_1 \int_0^1 dx_2
\frac{1}{\alpha_{{\bf p}}} \left\{1 -e^{-\alpha_{{\bf p}} t}\right\} =
\frac{6 k_B T}{\zeta} \int_0^t dt^\prime \int_0^1 dx_1 \int_0^1 dx_2
e^{-\alpha_{{\bf p}} t^\prime }.
\label{eq:msd6}
\end{equation}
We proceed with Eq. (\ref{eq:msd6}) by splitting the exponentials,
resulting in two identical integrals which can be evaluated in terms
of the modified Bessel function of the first kind $I_{\alpha}(z)$ and
making a change of variables.
\begin{equation}
\langle \Delta{\bf R}_{\text{m}}(t)^2 \rangle = \frac{6 k_B T}{\zeta}
\int_0^t dt^\prime \left( \int_0^1 dx \exp\left\{ - \frac{4 k
  t^\prime}{\zeta} \sin^2\left[ \frac{\pi}{2} x \right]\right\}
\right)^2= \frac{6 k_B T}{k} \int_0^{\frac{k t}{\zeta}} dt^\prime
e^{-4 t^\prime} I_0^2(2t^\prime).
\label{eq:msd9}
\end{equation}
To characterize the moderate time behavior of Eq. (\ref{eq:msd9}) we
look at the asymptotic expansion $I_{\alpha}(z)=\exp(z)/\sqrt{2\pi z}$
and evaluated the integral so that
\begin{equation}\langle \Delta{\bf R}_{\text{m}}(t)^2
  \rangle = \frac{3 k_B T}{2 \pi k} \left( \log\left[ \frac{k
      t}{\zeta} \right] + C_{\text{m}} \right),
\label{eq:msd10}
\end{equation}
where
\begin{equation}
C_{\text{m}} = \lim_{x\rightarrow\infty} -\log[x] + 4 \pi \int_0^x dt
e^{-4 t} I_0^2(2 t)\approx 4.04.
\label{eq:msd11}
\end{equation}

(b-c) The same can be done for the mean-square displacement for a bead
located in the middle of an edge $\langle \Delta{\bf
  R}_{\text{e}}(t)^2 \rangle$ and corner $\langle \Delta{\bf
  R}_{\text{c}}(t)^2 \rangle$. The corresponding calculations are
similar to those in (a), and they yield
\begin{equation}
\langle \Delta{\bf R}_{\text{e}}(t)^2 \rangle = \frac{3 k_B T}{\pi k}
\left( \log\left[ \frac{k t}{\zeta} \right] + C_{\text{e}} \right),
\label{eq:msd12}
\end{equation}
where
\begin{equation}
C_{\text{e}} = \lim_{x\rightarrow\infty} -\log[x] + 2 \pi \int_0^x dt
e^{-4 t} \left( I_{0}(2t) + I_{1}(2t) \right) I_0(2 t)\approx 2.47,
\label{eq:msd13}
\end{equation}
and
\begin{equation}
\langle \Delta{\bf R}_{\text{c}}(t)^2 \rangle = \frac{6 k_B T}{\pi k}
\left( \log\left[ \frac{k t}{\zeta} \right] + C_{\text{c}} \right),
\label{eq:msd14}
\end{equation}
where
\begin{equation}
C_{\text{c}} = \lim_{x\rightarrow\infty} -\log[x] + \pi \int_0^x dt
e^{-4 t} \left( I_{0}(2t) + I_{1}(2t) \right)^2 \approx 1.47.
\label{eq:msd15}
\end{equation}
\begin{figure}[h] \centering
\epsfig{file=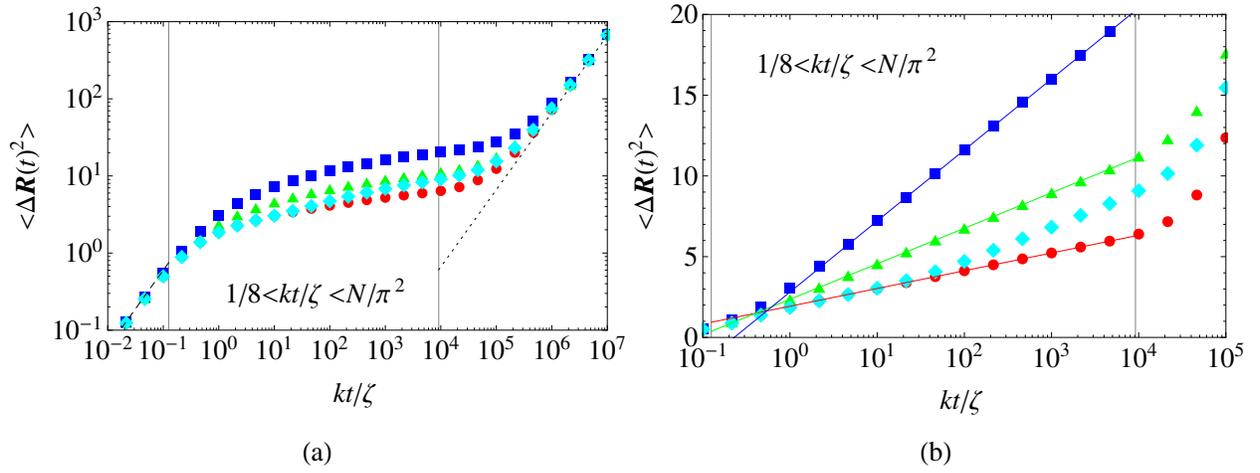,width=\linewidth,clip=}
\caption{The mean-squared displacement $\langle \Delta{\bf R}_{{\bf
      n}}(t)^2 \rangle$ for beads in a square membrane as a function
  of scaled time $k t / \zeta$. The sum in Eq. (\ref{eq:msd2}) was
  exactly evaluated for $k_BT/k=1$ for beads at the center of the
  membrane, at the middle of an edge, and at a corner for $N=301^2$
  and the result is represented by the circular (red), triangular
  (green), and square (blue) data points respectively. The cyan
  diamonds represent the exact evaluation for a bead with internal
  position ${\bf n}=(151,5)$ to show the transition that the
  mean-squared displacement for a bead makes from one region to
  another. For very short times $k t/\zeta\ll1/8$ the mean-squared
  displacement for all beads $\langle \Delta{\bf R}_{{\bf n}}(t)^2
  \rangle\sim t$ thus behaving like that of a free bead as shown in
  Eq. (\ref{eq:msd3}) which is represented by the dashed line in
  (a). For very long times the mean-squared displacement for all beads
  $\langle \Delta{\bf R}_{{\bf n}}(t)^2 \rangle\sim t/N$ behaves like
  that of a mean-squared displacement of the center-of-mass which is
  represented by the dotted line in (a). In the intermediate time
  regime the mean-square displacement for beads $\langle \Delta{\bf
    R}_{{\bf n}}(t)^2 \rangle\sim \log[k t/\zeta]$ as in agreement
  with Eqs. (\ref{eq:msd10}), (\ref{eq:msd12}) and (\ref{eq:msd14})
  --- the corresponding approximations are valid for $kt/\zeta\ll
  N/\pi^2$ --- and are represented by the red, green, and blue solid
  lines in (b) respectively.}
    \label{fig:msd}
\end{figure}

Note that all constants $C$'s can be evaluated up to arbitrary
precision, and that that the approximations (\ref{eq:msd9}),
(\ref{eq:msd11}) and (\ref{eq:msd13}) are valid for $kt/\zeta\ll
N/\pi^2$. In Fig. \ref{fig:msd} we compare the exact evaluations from
Eq. (\ref{eq:msd2}) and the corresponding approximations
(\ref{eq:msd3}), (\ref{eq:msd10}), (\ref{eq:msd12}) and
(\ref{eq:msd14}).

\subsection{Autocorrelation function of a vector connecting two beads}

In the most general case the vector ${\bf r}_{{\bf m}{\bf n}}(t)
\equiv {\bf R}_{{\bf m}}(t)-{\bf R}_{{\bf n}}(t)$ connects two beads
with internal coordinates ${\bf m}$ and ${\bf n}$ at time $t$. The
autocorrelation function of this vector $D_{{\bf m}{\bf n}}(t) \equiv
\langle {\bf r}_{{\bf m}{\bf n}}(t) \cdot {\bf r}_{{\bf m}{\bf n}}(0)
\rangle$ can be expressed in terms of the eigenmodes
\begin{equation}
D_{{\bf m}{\bf n}}(t) = \frac{3 k_B T}{\zeta N} \sum_{{\bf p} \neq
  {\bf 0}} \frac{\beta_{{\bf p}}}{\alpha_{{\bf p}}} \left[f_{{\bf
      p}}({\bf m}) - f_{{\bf p}}({\bf n})\right]^2
\exp\left[-\alpha_{{\bf p}} t\right].
\label{eq:acfsv1}
\end{equation}
Like in the case for the mean-square displacement of a tagged bead,
the behavior of $D_{{\bf m}{\bf n}}(t)$ depends on time and the
internal positions of the beads, hence we only briefly outline the
calculation procedure and state the results. We focus on two extreme
cases (a) $D_{\text{c}}(t)$ where the vector connecting the two beads
located at the opposite corners of the membrane, and (b)
$D_{\text{a}}(t)$ where the vector connects two beads at position
${\bf n}$ and ${\bf n}+{\bf a}$ somewhere in the middle of the
membrane for $a/L \equiv|{\bf a}|/L\ll1$. We will generalize these
results qualitatively for the cases when the locations of the beads
and the distances between them are arbitrary.

Short times: For times $kt/\zeta \ll 1/8$ the exponent is roughly $1$
and so the functions are constant. In case two beads are close
neighbors the sum can be evaluated $D_{\text{a}}(t)=k_B T (3/2+\log
a)/k$, which is exact for $a=1$, and a rough estimate
$D_{\text{c}}(t)=6 k_B T \log N/ (\pi k)$ follows from a calculation
similar to that performed for the radius of gyration.
\begin{figure}[!ht] \centering
\epsfig{file=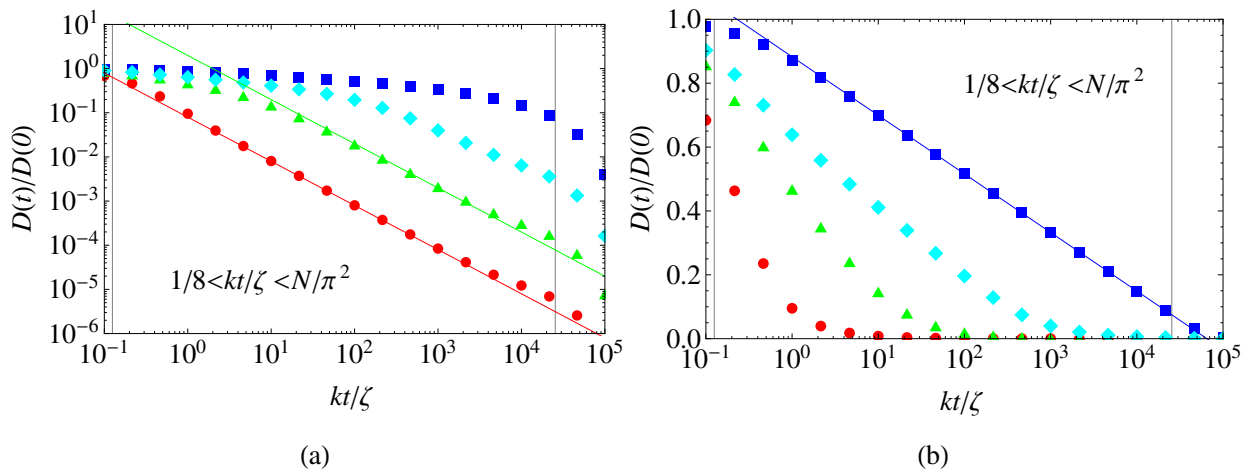,width=1.0\linewidth,clip=}
\caption{The scaled autocorrelation function $D_{{\bf m}{\bf
      n}}(t)/D_{{\bf m}{\bf n}}(0)$ of the vector connecting beads
  with internal position ${\bf m}$ and ${\bf n}$ in a square membrane
  as a function of scaled time $k t / \zeta$. The sum in
  Eq. (\ref{eq:acfsv1}) was exactly evaluated with $k_B T / k=1$ for
  two beads at positions [(251,251),(250,251)], [(251,254),(258,251)],
  [(1,1),(501,501)], and [(251,281),(281,251)] in the membrane with
  $N=501^2$, with the results shown by the circular (red), triangular
  (green), square (blue), and diamond (cyan) data points
  respectively. Note that the first two pairs of beads are very close
  to each other in the middle of the membrane where the third pair are
  two beads in opposing corners of the membrane; i.e., the first two
  cases correspond to $D_{\text{a}}(t)/D_{\text{a}}(0)$, while the
  third one corresponds to $D_{\text{c}}(t)/D_{\text{c}}(0)$. The
  function $D(t)\sim \log[k t/\zeta]$, as in agreement with
  Eq. (\ref{eq:acfsv2}) corresponding to the blue solid line in (b),
  for a duration that is larger when the beads are connected by a
  larger vector ${\bf a}$.  For very small vectors ${\bf a}$ the
  behavior will shift to $D(t)\sim 1/t$ as derived in
  Eq. (\ref{eq:acfsv3}) corresponding to the red and green solid lines
  in (a) before $kt/\zeta\sim N/\pi^2$. The function $D(t)$ for the
  last pair shows a transition from logarithmic behavior to $D(t)\sim
  1/t$ before the diffusive time regime.}
\label{fig:acfsv}
\end{figure}

Intermediate times: For $a/L\ll1$ the sum for $D_{\text{a}}(t)$ can be
expanded in $a$ and evaluated as
\begin{equation}
D_{\text{a}}(t)=\frac{3 \zeta k_BT}{8\pi k^2} \frac{a^2}{t}.
\label{eq:acfsv3}
\end{equation}
Further, it can be shown that the summation for $D_{\text{c}}(t)$
equals that of $-\langle \Delta{\bf R}_{\text{c}}(t)^2 \rangle$ up to
a constant, such that
\begin{equation}
D_{\text{c}}(t)=D_{\text{c}}(0)-\frac{6 k_B T}{\pi k} \left(
\log\left[ \frac{k t}{\zeta} \right] + C_{\text{c}} \right).
\label{eq:acfsv2}
\end{equation}

Long times: For $a/L\ll1$ and $k t/\zeta\gg N/\pi^2$ only the lowest
mode contributes to the summation and so expanding the correlation
function in Eq. (\ref{eq:acfsv1}) yields
\begin{equation}
D_{\text{a}}(t)=\frac{6 k_B T a^2}{k N} \exp\left[-\frac{\pi^2
    k}{\zeta N} t\right],
\label{eq:acfsv4}
\end{equation}
while
\begin{equation}
D_{\text{c}}(t)=\frac{48 k_B T}{\pi^2 k} \exp \left[ -\frac{\pi^2
    k}{\zeta N} t \right].
\label{eq:acfsv5}
\end{equation}

It is interesting to note the differences in behavior for
$D_{\text{a}}(t)$ and $D_{\text{c}}(t)$ at intermediate times. The
reason behind this difference is as follows. For $D_{\text{a}}(t)$ the
beads are close and as a result they quickly become `aware' of each
other's presence. For $D_{\text{c}}(t)$ on the other hand, the beads
do not become `aware' of each other's presence almost until
$kt/\zeta\sim N/\pi^2$.  Based on this observation we expect that when
the beads are neither very close nor very far, there will be first a
logarithmic decay (\ref{eq:acfsv2}) for $D_{{\bf m}{\bf n}}(t)$,
followed by a $1/t$ power-law decay (\ref{eq:acfsv3}) before the
terminal exponential decay (\ref{eq:acfsv4}-\ref{eq:acfsv5}) sets in.
The above results are verified by comparing the approximations to the
exact evaluation of Eq. (\ref{eq:acfsv1}) for different pairs of beads
in Fig. \ref{fig:acfsv}.

\section{Dynamical eigenmodes in the case of external tensile forces\label{sec4}}

In this section we concentrate on presenting the exact eigenmodes when
the membrane is stretched in two perpendicular directions by forces
${\bf F}_1$ and ${\bf F}_2$ applied at the edges of the membrane.

Adding tensile forces to the system ensures that each bead has its own
mean position around which it fluctuates due to thermal forces. For
large enough tensile forces the beads are far enough away from each
other, and self-intersection of the membrane is avoided. Such a
situation therefore mimics the behavior of a realistic flexible
membrane under tension. The system is still exactly solvable by
introducing the following term to the Hamiltonian
\begin{equation}
  U_F = \sum_{n_1,n_2=1}^{L_1-1,L_2} {\bf F}_1 \cdot \left( {\bf
    R}_{n_1,n_2} - {\bf R}_{n_1+1,n_2} \right)+
  \sum_{n_1,n_2=1}^{L_1,L_2-1} {\bf F}_2 \cdot \left( {\bf
    R}_{n_1,n_2} - {\bf R}_{n_1,n_2+1} \right),
\end{equation}
making the exercise of this section useful for practical purposes.

The mode amplitudes and their inverses are once again defined by
Eqs. (\ref{eq:eigenmode}-\ref{eq:inverse}), but the following term
${\bf H}_{{\bf p}}$ will be added to the right hand side of the
differential equation in Eq. (\ref{eq:eomeigenmode3}) in order to
solve for the dynamics of the mode amplitudes:
\begin{eqnarray}
{\bf H}_{{\bf p}}&=&\frac{1}{\zeta N} \sum_{{\bf n}} f_{{\bf p}}({\bf
  n}) \left[ {\bf F}_{1}(\delta_{n_1 L_1}-\delta_{n_1 1}) + {\bf
    F}_{2}(\delta_{n_2 L_2}-\delta_{n_2 1}) \right]\nonumber\\ &=&
\frac{1}{\zeta} \left[ \delta_{p_2 0} \{f_{p_1}(L_1)-f_{p_1}(1)\}
  \frac{{\bf F}_{1}}{L_{1}} + \delta_{p_1 0}
  \{f_{p_2}(L_2)-f_{p_2}(1)\} \frac{{\bf F}_{2}}{L_{2}} \right] .
\end{eqnarray}
Note that spatial symmetry for the modes is broken, and that
Eq. (\ref{eq:macfa}) is replaced by similar equations but also
depending on the spatial components $i$ and $j$ of the vectors.
\begin{equation}
\label{eq:macfa2}
X_{{\bf p} i {\bf q} j}(t) \equiv \langle {\bf X}_{{\bf p}i}(t) {\bf
  X}_{{\bf q}j}(0) \rangle = \frac{{\bf H}_{{\bf p}i}{\bf H}_{{\bf
      q}j}}{\alpha^2_{{\bf p}}} + \frac{k_B T}{\zeta N \beta_{{\bf
      p}}}\frac{1}{\alpha_{{\bf p}}} \exp \left[-\alpha_{{\bf p}}
  t\right] \delta_{{\bf p}{\bf q}} \delta_{i j} \quad \mbox{with }
{\bf p} \neq {\bf 0}.
\end{equation}
The assumption ${\bf F}_1\perp{\bf F}_2$ is necessary in order for the
modes to be the exact dynamical eigenmodes. If the tensile forces are
not orthogonal, then it will result in correlations between some of
the modes (and correspondingly a parallelogram-like structure of the
membrane).
\begin{figure}[h]
\centering \epsfig{file=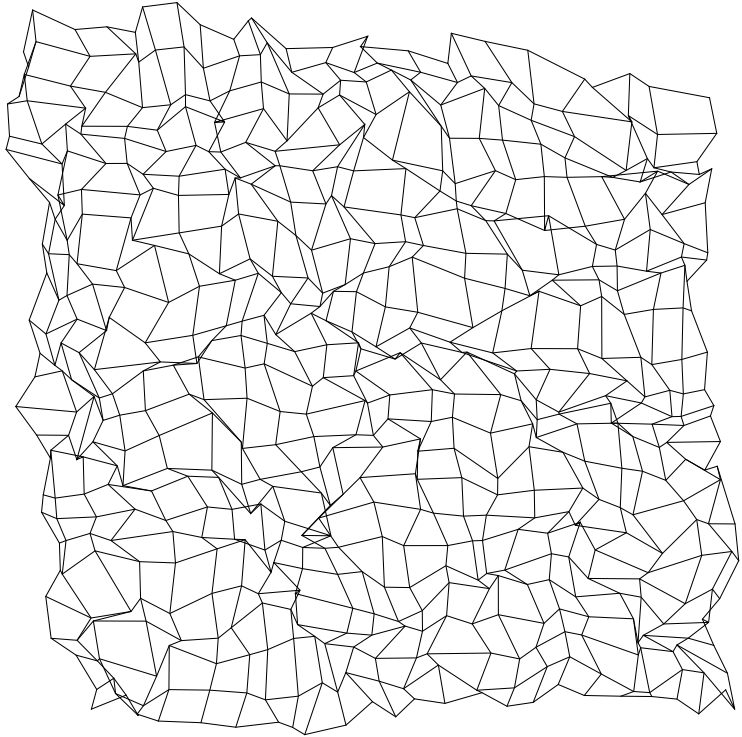,width=0.6\linewidth,clip=}
\caption{A randomly generated membrane consisting of $25\times25$
  beads in a three dimensional space, with orthonormal tensile forces,
  $k_B T=1/2$, and $k=\zeta=1$. The black lines connects neighboring
  beads. Due to tensile forces the membrane gets a rectangular lattice
  like structure where thermal forces push the beads out of their mean
  positions.}
\label{fig:membrane}
\end{figure}

It is of course interesting to investigate the transverse fluctuations
of the membrane under lateral tension. Two observables we consider are
the effective thickness and the additional surface area under thermal
undulations. A quantity that can be used as a measure for the
thickness of the membrane, is the standard deviation of the relative
height of the bead in the middle of the membrane with respect to its
mean position. In the limit of a continuous very large membrane, this
thickness is given by
\begin{equation}
D\equiv \sqrt{\left< R_{\text{m}}^2\right> }=\sqrt{\frac{k_B T}{4 k}}.
\label{neweq1}
\end{equation}
We verified numerically that the fluctuations increase for beads near
the edge of the membrane, in line with self-avoiding membranes
\cite{edge1,edge2}. Note that this thickness is unaffected by the
strength of the tensile forces.

A much more involved calculation is needed for the area of the
membrane. Consider a very large square membrane with $N=L^2$ beads and
perpendicular tensile forces $F_1=F_2=F$ of equal strength.  We consider
the case where the temperature is low enough, or the tensile forces
are strong enough, such that the fluctuations of the beads around their
equilibrium are small compared to the distances between two neighboring
beads.  The total area of the membrane can be written as the sum of areas
of all triangles between the three beads with indices $(i,j), (i+1,j)$
and $(i,j+1)$ plus the sum of areas of all triangles with indices
$(i+1,j), (i,j+1)$ and $(i+1,j+1)$; The expectation value for this area,
in the continuum limit, is then given by
\begin{equation}
\left< A \right>=\frac{F^2(L-1)^2}{k^2} + \frac{k_B T(2L^2-L)}{8 k}.
\label{neweq2}
\end{equation}

Equations (\ref{neweq1}-\ref{neweq2}) collectively imply that under
the application of tensile forces, irrespective of the force
magnitude, the out-of-plane degrees of freedom decouple from in plane
degrees of freedom. Moreover at strong stretching $F\gg k_BT/k$
the main contribution to the area of the membrane comes from the
stretching of the individual springs. As already mentioned earlier, at
strong stretching there is no difference between a phantom and a
self-avoiding membrane, and therefore we expect the same for the
increase in area for a self-avoiding membrane under strong stretching
forces.

The eigenmodes also give access to dynamical information. For instance,
to obtain the mean vector connecting two neighboring beads in the
membrane $\Delta {\bf R}_{1}(t) \equiv {\bf R}_{n_1+1,n_2}(t)-{\bf
  R}_{n_1,n_2}(t)$ with $n_1 = 1, \dots , L_1 - 1$ and $n_2 = 1,
\dots, L_2$, we use $\langle {\bf X}_{{\bf p}}(t) \rangle = {\bf
    H}_{{\bf p}}/\alpha_{{\bf p}}$, which can be proved by solving
the differential equation in a similar fashion as for
Eq. (\ref{eq:macfa2}), such that
\begin{equation}
\label{eq:equi}
\langle \Delta {\bf R}_{1}(t) \rangle = \sum_{{\bf p}} \beta_{{\bf p}}
\frac{{\bf H}_{{\bf p}}}{\alpha_{{\bf p}}} [f_{{\bf
      p}}(n_1+1,n_2)-f_{{\bf p}}(n_1,n_2)],
\end{equation}
yielding $\Delta {\bf R}_{1}(t)={\bf F}_{1}/k$. The same is true for
neighbors in the other internal direction so that for orthogonal
tensile forces, on average, the membrane obtains is a flat rectangular
structure. (The above results are obtained by using
\begin{equation}
\label{eq:identity}
\sum_{p=1}^{L-1} \sin^{-1}\left[\frac{\pi p}{2 L}\right]
\sin\left[\frac{\pi n p}{L}\right] \cos\left[\frac{\pi p}{2 L}\right]
= \frac{L}{2} \quad \mbox{for } n = 1,\dots, L-1 ,
\end{equation}
where the summation is over odd values of $p$). Further, the random
thermal forces are gaussian distributed with a variation given by
Eq. (\ref{eq:thermal}) such that $\langle {\bf X}^2_{{\bf
    p}i}(0)\rangle$ and $\langle {\bf X}_{{\bf p}}(0) \rangle$ can be
used to derive the first two moments for every mode amplitude. We use
these results to generate a random typical configuration for a
membrane as shown in Figure \ref{fig:membrane}.

\section{Discussion}\label{sec5}

The Rouse model has proved to be a cornerstone for polymer
dynamics. We generalized this model with two internal coordinates, for
which the entity becomes a phantom membrane. We have shown that a set
of eigenmodes can be used to solve the membrane dynamics in the
overdamped limit, and have demonstrated that many interesting
properties can be analytically derived for the membrane using certain
relations for these eigenmodes. Furthermore, we have shown that adding
large enough tensile forces to the system mimics the behavior of a
realistic flexible membrane under tension: the eigenmodes can still be
exactly solved analytically, making the analysis useful for practical
purposes.

We also note, although the following is not relevant to study the
properties of a phantom membrane, that the entire analysis presented
in this paper is easily generalized for any number $d\geq1$ of
internal coordinates. For every internal coordinate the Rouse modes
and inverse get a factor $f_{p_i}(m_i)$, $\beta_{{\bf p}}$ is a
product over $d$ factors, and $\alpha_{{\bf p}}$ becomes a summation
over $d$ terms. By generalizing the potential energy so that all
neighbors are connected, Eqs. (\ref{eq:ortho2}-\ref{eq:macfa}) remain
valid in their current forms. It can also be shown that adding a
tensile force for each internal coordinate still makes the system
analytically solvable. Similarly, the analysis can also be extended to
incorporate periodic boundaries (resulting in a torus or cylinder
geometry, as appropriate, when the internal dimension $d=2$).

\end{document}